# Hash Cracking Benchmarking of Replacement Patterns

Ensar Şeker
NATO CCD COE
ensar.seker@ccdcoe.org

*Abstract*— Enterprise password policies require the use of complex passwords that contain lowercase and uppercase letters, numbers and symbols. Considering this common requirement, end-users tend to create complex (!) passwords containing certain patterns which make such passwords guessable and therefore insecure. Replacement pattern is one of these pattern-types and substitutes a number or symbol for a certain letter. As an example, the letter "o" is replaced with 0 (zero) and password becomes passw0rd. Even though passw0rd contains a number and is assumed a strong password, its replacement pattern can be misused to guess it successfully and crack it easily. In our research, In our re-search, we performed an automated analysis of 1.330.780 word list from different languages to identify possible replacement patterns This list contains words form the dictionaries of the most used languages (only the ones that uses Latin script) in the world. We identified 43 different replacement-types for one character replacements (single type replacement), 9 different replacement-types for two character replacements (dual type replacement) and 15 different replacement-types for three character replacements (triad type replacement) for our analysis. These identified replacement patterns can be utilized to improve dictionary-attacks, especially for forensic investigations. In this paper, we explain our methodology to identify replacement patterns. The main purpose of this article is to show that with replacement methods on plain texts, it is possible to have more successful rates when trying to recovering hashed passwords.

## I. MOTIVATION

Authentication is one of the most important information security requirements and passwords are the most frequently used method for authentication due to its easy integration. But the security of passwords has always become and also will become a big headache for end-users, application developers, system administrators as well as security experts since black-hat hackers target primarily leakage of passwords [1, 2, 3, 4].

Passwords should be complex against guessing attacks. This requires that they should contain uppercase and lowercase letters, numbers, and symbols. Moreover, minimum password length should be 8 characters for today. But people as end-users, system administrators, etc. are bad at remembering complex passwords. It is also suggested that different passwords should be used for different server accounts and applications. This makes password security more problematic, even impossible to accomplish.

When people are forced to use complex passwords, they tend to choose passwords which fulfill complexity rules of policies whereas their chosen passwords are not secure against guessing attacks. As an example"Passw0rd." can be given. It contains an uppercase letter, lowercase letters, a number and a symbol. Even though it satisfies complexity rules according to the general complexity rules, it can be easily cracked by exploiting the patterns used. This password contains Capitalization pattern (i.e. p !P), Replacement pattern (i.e. o !0) as well as Appending pattern (i.e. a dot is appended).

In our research, we created plain text word list from variety of vocabularies. Then we applied different types of replacements to these words. For more accurate results, duplicated words were eliminated. We wrote a Python code for the replacements and eliminating duplicated results. Finally, we compared the result with and without replacements for recovering hashed passwords. Based on the results of our analysis, we identified 67 different replacement types. They can be utilized for example to enhance dictionary attacks. Especially, forensic investigators may need to bypass authentication for further analysis and they can benefit from our identified replacement patterns. In this paper, we explain our analysis methodology, our results in detail.

The paper is organized as follows: Section 2 discusses the related works. Our analysis of replacement patterns is explained in Section 3 in detail. The results of our analysis are evaluated in Section 4. Section 5 concludes the paper and discusses the future work.

## II. RELATED WORK

The first complexity analysis of RockYou passwords was conducted by Imperva[5] They focused only on complexity rather than patterns. According to their analysis, 60% of the RockYou passwords contain only lowercase letters, uppercase letters or numeric values and therefore very insecure. About 30% of the RockYou passwords have the length which is equal to or below six characters. The list of the most frequently used Top 20 passwords is given in the analysis as well. The most frequently used password is"123456".

Veras et al.[6] studied password patterns as well. They analyze numbers and different date formats in passwords rather than replacement patterns. Wu[7] performed pattern analysis as well as cracking tests of a Kerberos realm containing over 25000 users. In their guessing attack, they could crack a total of 2,045 passwords successfully by the end of a two-week experiment. The half of the successfully guessed passwords was cracked by conducting dictionary attack. For the remaining half, they utilized some patterns like prefix, suffix, capitalization and reversing.

Jakobsson and Dhiman[8] proposed a new password model based on an analysis of the RockYou passwords. They scored passwords from five datasets of disclosed password datasets (i.e. Rootkit, Sony, PayPal, Justin Bieber fan web page and Porn web page datasets). Their analysis shows the average number of components per password in the different datasets. As a result, they found out that Justin Bieber dataset has the highest average number of word components compared with the other analyzed datasets. Replacement pattern is one of the components they utilized, but their replacement types belong to a very small set of characters.

Wang et al.[9] proposed a new password strength meter which can detect also certain patterns and evaluate password security accordingly. Houshmand and Aggarwal[10] proposed a new system which analyzes whether a user proposed pass-word is weak or strong by estimating the probability of the password being cracked. They modify then the weak password to create a strengthened password as well. Some examples of the weak and strengthened password are trans2 !%trans2, colton00 !8colton00. This system is also insecure against pattern-based dictionary attacks. An attacker can delve into the details of this system, identify specific patterns used by this system and use these identified patterns to generate possible strengthened passwords.

Mazurek et al.[11] performed an empirical study of plain-text passwords of 25 thousand faculty, staff, and students at a research university. They concluded that certain groups create more secure passwords than the others. For example, computer science students make passwords more than 1.8 times as strong as the business school students. Comparing their contributions with our study, their focus is mainly based on the relation analysis of different categories like gender, college types, user types, etc. rather than password patterns.

Kelley et al.[12] studied the impact of different password policies on password complexity. They investigate mainly the resistance of passwords created under different policies and the performance of guessing algorithms under different training sets. Narayanan and Shmatikov[13] showed how to reduce the size of password search space for dictionary attacks by using Markov modeling techniques.

A new method was introduced by Matt Weir et al. [14]. They claimed that their method generates password structures in highest probability order. In light of a preparation set of previously disclosed passwords, they created a probabilistic context-free grammar in their method which can generate word-mangling rules. Based on their paper, this method was able to crack 28% to 129% more password than John the Ripper.

Cheng Yang, Jui-long Hung, and Zhangxi Lin[15] analyzed password patterns from a different perspective. Instead of general password patterns, they did their research on pass-word patterns of Chinese internet users. This research tried to understand how cultural factors effects molding password constructions. They reached some interesting results such as Chinese users have a weaker sense of security than Westerner users. The study of Hsien-Cheng Chou et al. [16] stated that they developed a password analysis platform which can analyzes commonly used passwords and identifies frequently used pass-word patterns. These analyses helped them to create a model consisting of a Training set, a Dictionary set, and a Testing set (TDT model) to generate probabilistic passwords arranged in diminishing request. They claimed TDT model successfully cracked more passwords 1.43 and 2.5 times higher than the John-the-Ripper attack and Brute-force attack.

Briland Hitaj et al.[17] involved deep learning approach for cracking password patterns for their research. Their machine learning technique (PassGAN) which can leverage Generative Adversarial Networks (GANs) to enhance password guessing. PassGAN represents a generous change in rule-based password generation tools since it induces password dissemination data self-rulingly from password information instead of by means of manual analysis. The paper compares their results with HashCat and underlines that PassGAN can match 18%-24% more passwords than HashCat alone.

There have been some studies in the past to evaluate password patterns as explained in this section. But most of these studies focus only on a very limited set of patterns. On the contrary, we developed a Python program that performed an automated analysis to identify all replacement patterns which people choose to create their strong(!) passwords. The advantage of the automated analysis is that it finds out a complete list of possible replacement types rather than a small set that can be found by manual analyzes.

## III. OUR ANALYSIS

### A. Creating A Word List

For this research, we used a plain text word list (1.330.780) from different languages including English, French, Spanish, Turkish, Italian, German, Dutch, Danish, and Norwegian which are the most common languages in the world that use Latin alphabets. All these dictionaries are available over the Internet. Table 1 illustrates the details about our word list.

| **Language** | **Number of Words** |
|---|---|
| English | 270.099 |
| French | 246.747 |
| Dutch | 180.130 |
| Spanish | 174.847 |
| German | 166.103 |
| Turkish | 119.575 |
| Italian | 88.351 |
| Norwegian | 61.413 |
| Danish | 23.515 |
| **Total** | **1.330.780** |

**Table 1** – Number of Worlds

The main reason to choose English, German, Spanish, French, Turkish, Italian, Dutch, Norwegian, and Danish for this article was that they are the most widely spoken languages which use Latin alphabet[18].

As it can be seen on Table 1, its English dictionary contains 270K words. Its French dictionary contains 240K words. Its Dutch dictionary contains 180K words. Its Spanish dictionary contains 174K words. Its German dictionary contains 166K words. Its Turkish dictionary contains 119K words. Its Italian dictionary contains 88K words. Its Norwegian dictionary contains 61K words. And finally, its Danish dictionary contains 23K words.

### B. Replacement Methodology

After creating a plain text list then we applied three types of combinations for replacements. First we used single letter combination using Table 2, then we used dual letters combination using Table 3, finally we used triad letters combination using Table 4. For dual letters and triad letters combinations we chose the replacement from the Top 5 List of Replacement Types (Table 5). So these replacements were chosen based on the statistics that shows most replaced characters.

| Nr | Replaced | Replacing | Examples |
|---|---|---|---|
| 1 | a | 0 | str0nger, fl0ppy, |
| 2 | a | 1 | sn1tch, ch1ppy |
| 3 | a | 4 | dr4gon, br4ndon |
| 4 | a | 8 | cre8tive, SK8TER |
| 5 | a | @ | Dr@gon, the@ter |
| 6 | b | 3 | num3ers |
| 7 | b | 6 | ra66it |
| 8 | b | 8 | bu88les, re8ecca |
| 9 | d | 0 | to0dles, no0dle |
| 10 | e | 0 | pr0view, cl0dle |
| 11 | e | 3 | str@wb3rry |
| 12 | e | 5 | mav5rick |
| 13 | e | 8 | dayl8ss |
| 14 | f | 4 | cali4ornia, the4ts |
| 15 | g | 6 | ma66ie, ti66er |
| 16 | g | 9 | ti99er |
| 17 | h | 1 | petler, nic1olas |
| 18 | h | 7 | mo7amed |
| 19 | i | 1 | pr1nc3ss, tr1n1ty |
| 20 | i | 7 | pr7ncess |
| 21 | i | 8 | sk8er, sk8ing |
| 22 | i | ! | jess!ca, Pr!nc3ss |
| 23 | l | 1 | welc0me, app1es |
| 24 | l | 7 | me7issa |
| 25 | l | ; | che;sea ho;;ywood |
| 26 | l | ! | vo!!ey |
| 27 | m | , | ja,es |
| 28 | o | 0 | pe0ple, vict0ria |
| 29 | o | 3 | ch33se |
| 30 | o | @ | passw@rd |
| 31 | r | . | st.anger, st.oh |
| 32 | s | 1 | pa11word |
| 33 | s | 2 | pa22word |
| 34 | s | 3 | pa33word, |
| 35 | s | 4 | pas4word |
| 36 | s | 5 | ca55ie, mon5ter |
| 37 | s | 6 | pa66word |
| 38 | s | 8 | pa88word |
| 39 | s | $ | je$$ica, Pa$$word |
| 40 | t | 7 | Ma77hew |
| 41 | t | 8 | la8ter |
| 42 | v | 7 | Se7en |
| 43 | z | ? | gon?alo |

**Table 2** – Single Type Replacement

| Nr | Replaced 1 | Replacing 1 | Replaced 2 | Replacing 2 |
|---|---|---|---|---|
| 1 | a | @ | o | 0 |
| 2 | a | @ | i | 1 |
| 3 | a | @ | l | 1 |
| 4 | a | @ | e | 3 |
| 5 | i | 1 | o | 0 |
| 6 | i | 1 | e | 3 |
| 7 | o | 0 | e | 3 |
| 8 | o | 0 | l | 1 |
| 9 | l | 1 | e | 3 |

**Table 3** – Dual Type Replacement

| Nr | Replaced 1 | Replacing 1 | Replaced 2 | Replacing 2 | Replaced 3 | Replacing 3 |
|---|---|---|---|---|---|---|
| 1 | a | @ | o | 0 | i | 1 |
| 2 | a | @ | o | 0 | l | 1 |
| 3 | a | @ | o | 0 | e | 3 |
| 4 | a | @ | l | 1 | e | 3 |
| 5 | a | @ | i | 1 | e | 3 |
| 6 | i | 1 | o | 0 | e | 3 |
| 7 | l | 1 | o | 0 | e | 3 |
| 8 | s | $ | l | ! | o | @ |
| 9 | s | $ | i | ! | o | @ |
| 10 | s | $ | l | ! | a | @ |
| 11 | s | $ | i | ! | a | @ |
| 12 | b | 6 | g | 9 | l | 1 |
| 13 | b | 6 | g | 9 | s | 5 |
| 14 | g | 9 | l | 1 | s | 5 |
| 15 | b | 6 | l | 1 | s | 5 |

**Table 4** – Triad Type Replacement

| Nr | Replaced | Replacing | Example |
|---|---|---|---|
| 1 | i | 1 | mon1ka (monika), cook1es (cookies), fall1ng (falling) |
| 2 | o | 0 | pe0ple (people), ver0nica (veronica), mem0ries (memories) |
| 3 | e | 3 | hat3rs, spid3rman (spiderman), st3lla (stella) |
| 4 | l | 1 | car1os (carlos), wa11ace (wallece) |
| 5 | a | @ | tiff@ny (tiffany), sp@rky (sparky) |

**Table 5** – Top 5 List of Replacement Types

### C. Recovering Hashed Passwords

For recovering hashed passwords, hashcat was preffered for better performance and more accurate results. Just like creating a plain text database, to measure success rate we created a hashed passwords. The total number of hahshed passwords in this database is 1.411.217. The details of hashed passwords are in Table 6. Duplicated hashes were removed the database so after the elimination 825.211 unique input remain.

| Name | Number of Hashes |
|---|---|
| dhool | 15.301 |
| gamish | 50.853 |
| sunrise | 7.660 |
| ffgbeach | 481.377 |
| OpNKorea | 9.001 |
| bfield | 548.686 |
| rootkit | 71.228 |
| whitefox | 47.238 |
| dsl | 14.144 |
| opisrael | 10.809 |
| mayhem | 130.884 |
| casio | 24.035 |
| **Total** | **1.411.217** |
| **Total (after removing duplicates)** | **825.211** |

**Table 6** – Number of Hashes

### D. Results

As mentioned before, we created dictionary inputs which consist of words from different languages. The total number of the word is 1.330.780. With this word list, we try to recover 825.211 unique MD5 hashes and 17.210 of them were successfully recovered. Then by applying our 3 different kind of pattern replacement to 1.330.780, we created 33.171.129 words as a new input. This time, with this world list, we try to recover 825.211 unique MD5 hashes again and 30.215 of them were successfully recovered. As a result, with replacement patterns success rate of recovering hashed password considerably higher. When we compare both results, recovering passwords with patterns is almost 80 percent more higher then recovering them with just regular words.

It is obvious from the results of our analysis that replacement patterns are commonly used by end-users and they jeopardize password security and authentication processes. The following mitigation methods can be suggested to enhance password security.

Secure password managers (SPM) can be utilized to generate strong passwords without any pattern and store them securely by using encryption methods (e.g. AES-256). End-users need to create a strong master password to access their SPM database. Some SPMs support even smart-card authentication or/and two-factor authentication by checking the existence of a physical le that is generated randomly during the setup phase.

Considering today's risk scenarios, authentication based only on passwords is a very insecure method especially for critical applications like email, online banking, social networking, etc. Two-factor authentication should be activated and utilized especially for such critical applications. As examples, most online banking portals, Google Mail, Twitter, Wordpress etc. support already software tokens that are sent over SMS or generated by a native mobile application (e.g. Google Authenticator) for two-factor authentication.

Awareness about patterns should be improved as well. Information security training organized especially for non-security experts can take patterns into consideration and ex-plain the risks of password patterns in detail. Black-listing of passwords can be extended by integrating patterns and not accepting pattern-based passwords by authentication systems.

IV. CONCLUSION AND FUTURE WORK

End-users need to choose complex passwords consisting of uppercase and lowercase letters, numbers, and symbols in order to prevent password guessing attacks. Since randomly-generated complex passwords are difficult to remember, they tend to create pattern-including passwords which contain numbers and symbols but are non-resistant to guessing at-tacks. In our research, we focused on a special type of such patterns namely Replacement Pattern. We identified possible replacement-types and try to recover passwords successfully. The identified replacement-types can be used to improve guessing attacks, especially for forensic investigations. Secure pass-word managers and two-factor authentication are suggested to minimize the risks of replacement patterns.

As future work, we plan to improve automated analyzes of other patterns like Inserting, Repeating, Sequencing and Capitalizing can be performed.